\documentclass[aps,preprint,amsmath,amssymb]{revtex4}
\usepackage{graphicx}
\def\be{\begin{eqnarray}}
\def\en{\end{eqnarray}}
\def\non{\nonumber}

\def\pl{{ Phys. Lett.}~}
\def\pr{{ Phys. Rev.}~}
\def\prl{{ Phys. Rev. Lett.}~}
\def\bi{\bibitem}
\begin{document}

\title{\Large \bf Model-independent analysis for determining mass splittings of heavy baryons
 }

\author{ \bf \large Chien-Wen Hwang\footnote{Email: t2732@nknucc.nknu.edu.tw}
 }

\affiliation{\centerline{Department of Physics, National Kaohsiung Normal University,} \\
\centerline{Kaohsiung, Taiwan 824, Republic of China}
 }


\begin{abstract}
We study the hyperfine mass differences of heavy hadrons in the heavy quark effect theory (HQET). The
effects of one-gluon exchange interaction are considered for the heavy mesons and baryons. Base on
the known experimental data, we predict the masses of some heavy baryons in a model-independent way.
\end{abstract}
\maketitle %

\section{Introduction}
It is widely accepted that Quantum Chromodynamics (QCD) is the
correct theory for strong interactions. QCD is a renormalizable
quantum field  theory which is closely modeled after quantum
electrodynamics (QED), the most accurate physical theory we have to
date. However, in the low-energy regime, QCD tells us that the
interactions between quarks and gluons are strong, so that
quark-gluon dynamics becomes non-perturbative in nature.
Understanding the structures of hadrons directly from QCD remains an
outstanding problem, and there is no indication that it will be
solved in the foreseeable future. In 1989, it was realized that, in
low energy situations where the typical gluon momenta are small
compared with the heavy quark mass $(m_Q)$, QCD dynamics becomes
independent of the flavor and spin of the heavy quark
\cite{iw,Geo1}. For the heavy flavors, this new symmetry called
heavy quark symmetry (HQS). Of course, even in this infinite heavy
quark mass limit, low energy QCD dynamics remains non-perturbative,
and what HQS can do for us is to relate otherwise unrelated static
and transition properties of heavy hadrons, and hence enormously
reduces the complexity of theoretical analysis. In other words, HQS
allows us to factorize the complicated light quark and gluon
dynamics from that of the heavy one, and thus provides a clearer
physical picture in the study of heavy quark physics. Beyond the
symmetry limit, a heavy quark effective theory (HQET) can be
developed by systematically expanding the QCD Lagrangian in powers
of $1/m_Q$, with which HQS breaking effects can be studied order by
order \cite{Geo1,NN,MRR}.

In the experimental area, all masses of $s$-wave charmed hadrons and
bottomed mesons which containing one heavy quark are found at
present. However, except the particle $\Lambda_b^0$ was already
found in the early 1980's, there has not been significant progress
in searching $s$-wave bottomed baryons until these months. Recently
some bottomed baryons were discovered at Fermilab. They are the
exotic relatives of the proton and neutron $\Sigma^{(*)+}_b$ and
$\Sigma^{(*)-}_b$ by CDF collaboration \cite{CDF} and the
triple-scoop baryon $\Xi_b^-$ by D0 and CDF collaborations
\cite{DZero,CDF1}. It is reasonable that the remainder particles,
which include $\Xi'_b$, $\Xi_b^*$, $\Omega_b$, and $\Omega_b^*$,
will be observed in the foreseeable future. All these heavy hadrons
provide a testing ground for HQET with the phenomenological models
to the low-energy regime of QCD. In this paper we focus on one
static property, that is, the mass spectrum of heavy hadrons and
combine HQET with the known experimental data to predict the mass
splitting of some heavy baryons. The phenomenological models are not
needed here.

The paper is organized as follows. In Sec. II brief introductory notes are given for HQET. In Sec.
III we formulate the hyperfine mass splitting for heavy mesons and baryons. In Sec. IV we evaluate
the numerical results and predict some mass differences between heavy baryons. Finally, the
conclusion is given in Sec. V.
\section{Heavy quark effect theory}

The full QCD Lagrangian for a heavy quark ($c$, $b$, or $t$) is given by
 \be
   {\cal L}_Q = \bar Q~(i\gamma_\mu D^\mu - m_Q)~Q,   \label{Lag}
 \en
where $D^\mu \equiv \partial ^\mu - i g_s T^a A^{a\mu}$ with $T^a = \lambda^a/2$. Inside a hadronic
bound state containing a heavy quark, the heavy quark $Q$ interacts with the light degrees of freedom
by exchanging gluons with momenta of order $\Lambda_{QCD}$, which is much smaller than its mass
$m_Q$. Consequently, the heavy quark is close to its mass shell, and its velocity does not deviate
much from the hadron's four-velocity $v$. In other words, the heavy quark's momentum $p_Q$ is close
to the ``kinetic" momentum $m_Q v$ resulting from the hadron's motion
 \be
   p^\mu_Q = m_Q v^\mu + k^\mu,  \label{Pk}
 \en
where $k^\mu$ is the so-called ``residual" momentum and is of order
$\Lambda_{QCD}$. To describe the properties of such a system which
contains a very heavy quark, it is appropriate to consider the limit
$m_Q \rightarrow \infty$ with $v$ and $k$ being kept fixed. In this
limit, it is evident that the quantity $m_Q v$ is ``frozen out" from
the QCD dynamics, so it is appropriate to introduce the ``large" and
``small" component fields $h_v$ and $H_v$, which is related to the
original field $Q(x)$ by
 \be
   h_v (x) = e^{im_Q v\cdot x} P_+ Q(x),  \non \\
   H_v (x) = e^{im_Q v\cdot x} P_- Q(x),
 \en
where $P_+$ and $P_-$ are the positive and negative energy projection operators respectively: $P_\pm
= (1 \pm \not\!v)/2$. so that
 \be
   Q (x) = e^{-im_Q v \cdot x} \left[h_v(x) + H_v (x) \right].  \label{hH}
 \en
It is clear that $h_v$ annihilates a heavy quark with velocity $v$,
while $H_v$ creates a heavy antiquark with velocity $v$. In the
heavy hadron's rest frame $v=(1,\vec 0)$, $h_v(H_v)$ correspond to
the upper (lower) two components of $Q (x)$.

   In terms of the new fields, the QCD Lagrangian for a heavy quark given by Eq. (\ref{Lag}) takes the following form
 \be
   {\cal L}_Q = \bar h_v i v\cdot D h_v - \bar H_v (i v\cdot D + 2 m_Q) H_v + \bar h_v i\not\!\!D_{\bot} H_v +
   \bar H_v i\not\!\!D_{\bot} h_v  \label{nLag}
 \en
where $D^\mu_{\bot} = D^\mu - v^\mu v \cdot D$ is orthogonal to the heavy quark velocity, $v\cdot
D_{\bot} = 0$. From Eq. (\ref{nLag}), we see that $h_v$ describes massless degrees of freedom,
whereas $H_v$ corresponds to fluctuations with twice the heavy quark mass. The heavy degrees of
freedom represented by $H_v$ can be eliminated using the equations of motion of QCD. Substituting Eq.
(\ref{hH}) into $ (i\not\!\!D - m_Q) Q (x) = 0$ gives
 \be
   i\not\!\!D h_v + (i\not\!\!D - 2 m_Q) H_v = 0.
 \en
Multiplying this equation by ${\cal P}_\pm$, one obtains
 \be
   -i v \cdot D h_v = i \not\!\!D_\bot H_v,  \label{1-14}  \\
   (i v \cdot D + 2 m_Q) H_v = i \not\!\!D_\bot h_v. \label{1-15}
 \en
The second equation can be solved schematically to give
 \be
   H_v = {1\over {(i v \cdot D + 2 m_Q -i\epsilon)}} i \not\!\!D_\bot h_v,
 \en
which shows that the small component field $H_v$ is indeed of order $1/m_Q$. One can insert this
solution back into Eq. (\ref{1-14}) to obtain the equation of motion for $h_v$. It is easy to check
that the resulting equation follows from the effective Lagrangian
 \be
   {\cal L}_{Q,eff} = \bar h_v i v\cdot D h_v + \bar h_v i \not\!\!D_{\bot} {1\over {(i v \cdot\! D +
   2 m_Q -i\epsilon)}} i \not\!\!D_\bot h_v.  \label{Lm}
 \en
${\cal L}_{Q,eff}$ is the Lagrangian of the heavy quark effective theory (HQET), and the second term
of Eq. (\ref{Lm}) allows for a systematic expansion in terms of $i D/m_Q$. Taking into account that
$P_+ h_v = h_v$, and using the identity
 \be
   P_+ i \not\!\!D_{\bot} i \not\!\!D_\bot P_+ = P_+ \left[(iD_\bot)^2 + {g_s \over {2}}
   \sigma_{\alpha\beta} G^{\alpha\beta}\right] P_+,
 \en
where
 \be
   G^{\alpha\beta} = T_a G^{\alpha\beta}_a = {i\over {g_s}}[D^\alpha,D^\beta]
 \en
is the gluon field strength tensor, one finds that
 \be
   {\cal L}_{Q,eff} = \bar h_v i v\cdot D h_v + {1\over {2 m_Q}} \bar h_v (iD_\bot)^2 h_v + {g\over {4 m_Q}}
   \bar h_v \sigma_{\alpha\beta} G^{\alpha\beta} h_v + {\cal O} ({1\over {m^2_Q}}).     \label{expand}
 \en
The new operators at order $1/m_Q$ are
 \be
   {\cal O}_1 = {1\over {2 m_Q}}~{\bar h}_v~(iD_\bot)^2~ h_v,  \label{O1} \\
  {\cal O}_2 = {g\over {4 m_Q}}~{\bar h}_v~\sigma^{\mu\nu}~G_{\mu\nu}~h_v, \label{O2}
 \en
where ${\cal O}_1$ is the gauge invariant extension of the kinetic
energy arising from the off-shell residual motion of the heavy
quark, and ${\cal O}_2$ describes the color magnetic interaction of
the heavy quark spin with the gluon field. It is clear that both
${\cal O}_1$ and ${\cal O}_2$ break the flavor symmetry, while
${\cal O}_2$ breaks the spin symmetry as well. For instance, ${\cal
O}_1$ would introduce a common shift to the masses of pseudoscalar
and vector heavy mesons, and ${\cal O}_2$ is responsible for the
color hyperfine mass splittings $\delta m_{_{HF}}$.

The full expansion of  ${\cal L}_{Q,eff}$ in $i D/m_Q$ can be organized as follow
 \be
   {\cal L}_{Q,eff} = \sum_{n=0} \left({1\over {2 m_Q}}\right)^n {\cal L}_n \label{expansion}
 \en
where the first few terms are given by
 \be
   {\cal L}_0 &=& \bar h_v (i~v \cdot D) h_v,  \non \\
   {\cal L}_1 &=& \bar h_v (iD_\bot)^2 h_v + {g\over {2}} \bar h_v \sigma_{\alpha\beta} G^{\alpha\beta} h_v,
   \non \\
   {\cal L}_2 &=& g \bar h_v \sigma_{\alpha\beta} v_\gamma i D^\alpha G^{\beta\gamma} h_v + g \bar h_v v_\alpha
   i D_\beta G^{\alpha\beta} h_v. \non
 \en
We must emphasize that this effective theory comes from first
principle directly, and it is in terms of the power of $1/m_Q$,
which is small enough, to calculate the physical quantities which
concerning heavy quarks perturbatively, i.e., order by order. If one
chooses the appropriate frame and phenomenological model, then one
can handle many physical processes systematically. We also note
that, as shown in Eq. (\ref{expansion}), the information of heavy
quark flavor is only involved in the factor $(1/2m_Q)^n$. In other
words, all ${\cal L}_n$'s are independent of the heavy quark flavor.
We will use this property to evaluate the mass splittings of some
heavy hadrons.

\section{Hyperfine mass splitting}
First, we consider the hyperfine mass splitting between pseudoscalar and vector heavy mesons. The
operators that break HQS to order $1/m_Q$ are ${\cal O}_1$ and ${\cal O}_2$ given in Eqs. (\ref{O1})
and (\ref{O2}) respectively. ${\cal O}_1$ can be separated into a kinetic energy piece and a
one-gluon exchange piece:
 \be
   {\cal O}_1 = {\cal O}_{1k} + {\cal O}_{1g}
 \en
where
 \be
   {\cal O}_{1k} &\equiv& {-1\over {2m_Q}} \bar h_v [\partial_\mu \partial^\mu+ (v\cdot \partial)^2]h_v,  \\
   {\cal O}_{1g} &\equiv& {-g_s\over {2m_Q}} \bar h_v [(p+p')_\mu - v\cdot (p+p')v_\mu]A^\mu~h_v,
 \en
Also, ${\cal O}_2$ can be reexpressed as
 \be
   {\cal O}_2 = -g_s T^a \sigma_{\mu\nu} \partial^\mu A^{a\nu}.
 \en
With the $1/m_Q$ corrections included, the heavy meson masses can be expressed as
 \be
   M_M = m_Q + {\bar \Lambda}^q -{1\over {2 m_Q}}(\lambda_1^q + d_{M} \lambda^q_2),\label{MMM}
 \en
where $\lambda^q_{1}$ comes from ${\cal O}_1$ and $\lambda_{2}^q$
comes from ${\cal O}_2$. $\lambda_{1}^q$ receive two different
contributions, one from ${\cal O}_{1k}$ and the other from ${\cal
O}_{1g}$, thus
 \be
   \lambda_{1}^q = \lambda_{1k} + \lambda_{1g}^q.
 \en
The parameter ${\bar \Lambda}^q$ in Eq. (\ref{MMM}) is the residual
mass of heavy mesons in the heavy quark limit. In other words,
${\bar \Lambda}^q$ is independent of heavy quark flavor.
$\lambda_{1k}$ comes from the heavy quark kinetic energy,
$\lambda_{1g}^q$ and $\lambda_{2}^q$ are respectively the
chromoelectric and chromomagnetic contributions. $\lambda_{1}^q$
parameterizes the common mass shift for the pseudoscalar and vector
mesons, and $\lambda_{2}^q$ accounts for the hyperfine mass
splitting. In both the non-relativistic and relativistic quark
models, the hyperfine mass splitting comes from a spin-spin
interaction of the form
 \be
   H_{HF} \sim \vec S_Q \cdot \vec J_l.
 \en
where $\vec S_Q$ is the spin operator of the heavy quark and $\vec
J_l$ is the angular momentum operator of the light degree of
freedom. Thus
 \be
   d_{M} &=& -\langle M(v) |4 \vec S_Q \cdot \vec J_l |M(v)\rangle \non \\
&=& -2[S_M(S_M+1) - S_Q(S_Q+1)- J_l(J_l+1)],
 \en
where $S_M$ is spin quantum number of the meson $M$. Consequently,
$d_{M} = -1$ for a vector meson, and $d_{M} = 3$ for a pseudoscalar
meson. Therefore, we obtain the hyperfine mass splitting,
 \be
  \Delta M_{VP} \equiv M_V-M_P = {2 \lambda_{2}^q\over {m_Q}}. \label{masslp}
 \en

We next consider the hyperfine mass splitting among the baryons
containing one heavy quark ($Q$) and two light quarks ($q_1$,
$q_2$). Each light quark is in a triplet $q=(u,d,s)$ of the flavor
$SU(3)$. Since $3\otimes 3 = 6\oplus\bar 3$ and the lowest lying
light quark state has $n=1$ and $L=0$ ($S$-wave), there are two
different diquarks: a symmetric sextet ($\vec{J}_l=1$) and an
antisymmetric antitriplet ($\vec{J}_l=0$). When the diquark combines
with a heavy quark, the sextet contains both spin-$\frac{1}{2}$
(${\cal B}_6$) and spin-$\frac{3}{2}$ (${\cal B}^*_6$) baryons, and
the antitriplet contains only spin-$\frac{1}{2}$ (${\cal B}_{\bar
3}$) baryons. The multiplets ${\cal B}_{\bar 3}$ and ${\cal
B}_6^{(*)}$ are illustrated in Fig. 1 (a) and (b), respectively, and
their quantum numbers are listed in TABLE I.
\begin{figure}[hb]
\hskip 8.5cm
~~~~\includegraphics[%
  scale=0.7,
  angle=0]{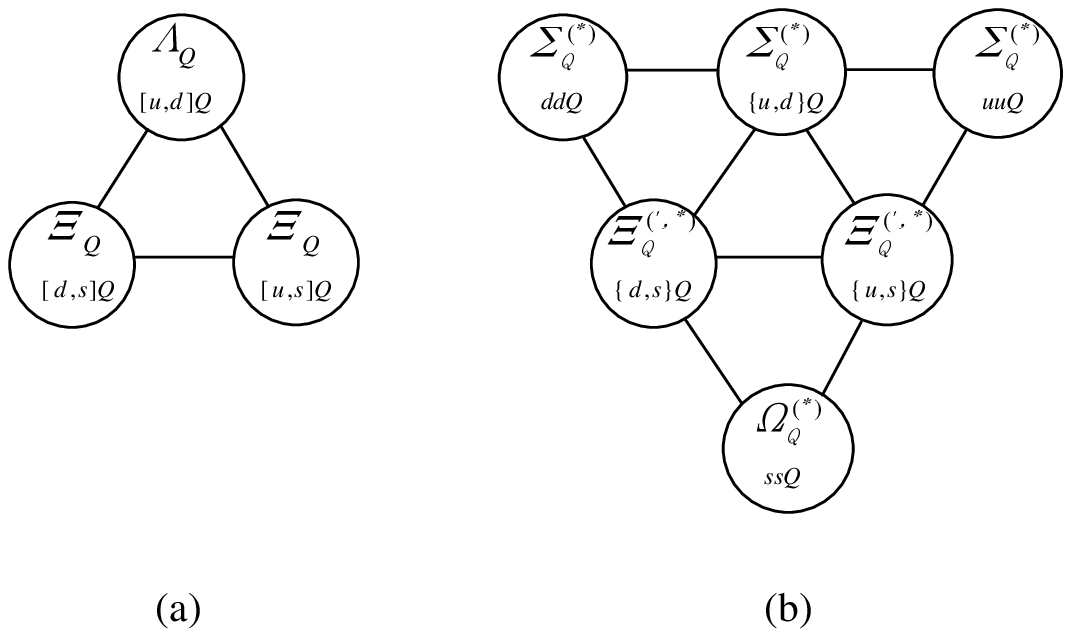}
\caption{The multiplets (a) $B_{\bar 3}$ and (b) $B_6^{(*)}$.}
\end{figure}

\begin{table}
\caption{\label{tab1} The $s$-wave heavy baryons and their quantum
number, where the subscript $l$ stands for the quantum number of the
two light quarks.}
\begin{ruledtabular}
\begin{tabular}{c|ccc|ccc|cc}
 state  & $\Lambda_Q$ & $\Sigma_Q$ & $\Sigma_Q^*$ & $\Xi_Q$ & $\Xi'_Q$ & $\Xi_Q^*$
 & $\Omega_Q$ & $\Omega_Q^*$ \\ \hline
 $J^P$  & ${1\over {2}}^+$ & ${1\over {2}}^+$ & ${3\over {2}}^+$ & ${1\over {2}}^+$ &
 ${1\over {2}}^+$ & ${3\over {2}}^+$ & ${1\over {2}}^+$ & ${3\over {2}}^+$  \\
 $J_l$  & $0$ & $1$ & $1$ &$0$ &$1$ &$1$ &$1$ &$1$  \\
\end{tabular}
\end{ruledtabular}
\end{table}
By analogy with Eq. (\ref{MMM}), the heavy baryon masses can be
expressed as
 \be
   M_{\cal B} = m_Q + {\bar \Lambda}_{J_l}^{q_1 q_2} -{1\over {2
   m_Q}}
   (\lambda^{q_1q_2}_1 + d_{\cal B} \lambda^{q_1q_2}_2), \label{MMB}
 \en
where ${\bar \Lambda}^{q_1 q_2}_{J_l}$ is the residual mass of heavy
baryons in the heavy quark limit. The proportions of
$\lambda^q_{1,2}$ (for meson) to $\lambda^{q_1 q_2}_{1,2}$ (for
baryon) are \cite{Jenkins}
 \be
 \lambda_{1}^q &\sim& \lambda_{1}^{q_1 q_2}, \non \\
 \lambda_{2}^q &\sim& N_c \lambda^{q_1 q_2}_{2},
 \en
where $N_c$ is the color number. Thus
 \be
 d_{\cal B}&=& -\langle {\cal B}(v) |4 (\vec S_Q \cdot \vec J_{l})
 |{\cal B}(v)\rangle \non \\
 &=& -2[S_{\cal B}(S_{\cal B}+1) - S_Q(S_Q+1) - J_l(J_l+1)],
 \en
where $S_{\cal B}$ is spin quantum number of the baryon ${\cal B}$.
Consequently, $d_{\cal B} = 0$ for a ${\cal B}_{\bar 3}$ baryon,
$d_{\cal B} = 4$ for a ${\cal B}_6$ baryon, and $d_{\cal B} = -2$
for a ${\cal B}_6^*$ baryon. Therefore, the hyperfine mass
splittings of $\Lambda_Q$, $\Sigma_Q$, and $\Sigma_Q^*$ are
 \be
 \Delta M_{\Sigma_Q^* \Sigma_Q}&=& {3 \lambda^{\tilde{q}_1 \tilde{q}_2}_{2}\over
 {m_Q}},\label{Sigma6s6} \\
 \Delta M_{\Sigma_Q \Lambda_Q} &=& {-2 \lambda^{\tilde{q}_1 \tilde{q}_2}_{2}\over {m_Q}}+
 \delta{\bar \Lambda}^{\tilde{q}_1\tilde{q}_2}, \label{Sigma63} \\
 \Delta M_{\Sigma_Q^* \Lambda_Q} &=& { \lambda^{\tilde{q}_1 \tilde{q}_2}_{2}\over {m_Q}}+
 \delta{\bar \Lambda}^{\tilde{q}_1\tilde{q}_2}, \label{Sigma6s3}
 \en
where $\tilde{q}$ is the $u$ or $d$ quark and $\delta{\bar
\Lambda}^{q_1 q_2}={\bar \Lambda}^{q_1 q_2}_1-{\bar \Lambda}^{q_1
q_2}_0$. For the $\Xi_Q$ baryons, however, the complexity is
increased because of the flavor $SU(3)$ symmetry breaking. We
consider the flavor $SU(3)$ symmetry breaking and write down the
hyperfine mass splittings of $\Xi_Q$, $\Xi'_Q$, and $\Xi_Q^*$ as
 \be
 \Delta M_{\Xi_Q^* \Xi'_Q} &=& {3 \lambda^{s\tilde{q}}_{2}\over
 {m_Q}},\label{Xi6s6} \\
 \Delta M_{\Xi'_Q \Xi_Q} &=& {-2 \lambda^{s\tilde{q}}_{2}\over {m_Q}}+
 \delta{\bar \Lambda}^{s\tilde{q} }, \label{Xi63} \\
 \Delta M_{\Xi_Q^* \Xi_Q} &=& { \lambda^{s \tilde{q}}_{2}\over {m_Q}}+
 \delta{\bar \Lambda}^{s \tilde{q}}. \label{Xi6s3}
 \en
It is worth to mention that, from Eqs. (\ref{Sigma63}),
(\ref{Sigma6s3}), (\ref{Xi63}), and (\ref{Xi6s3}), $\delta{\bar
\Lambda}^{q_1 q_2}$ are
 \be
 \delta{\bar\Lambda}^{\tilde{q}_1
 \tilde{q}_2}&=&{M_{\Sigma_Q}+2M_{\Sigma^*_Q}\over{3}}-M_{\Lambda_Q}, \label{Sigmaspinave}\\
 \delta{\bar\Lambda}^{s\tilde{q}}&=&{M_{\Xi'_Q}+2M_{\Xi^*_Q}\over{3}}-M_{\Xi_Q} \label{Xispinave}.
 \en

Finally, the hyperfine mass splitting of $\Omega_Q$ and $\Omega_Q^*$
is
 \be
 \Delta M_{\Omega_Q^* \Omega_Q} &=& {3 \lambda^{ss}_{2}\over
 {m_Q}}.\label{Omega6s6}
 \en
We can use the hyperfine mass differences which are experimentally known for charmed baryons to
calculate the ones for bottomed baryons.
\section{Numerical results}
Now we consider the numerical results of the hyperfine mass splitting for heavy mesons. As mentioned
above, $\lambda_{2q}$ just relates to the light degrees of freedom and independent of the heavy quark
mass $m_Q$. Thus, from Eq. (\ref{masslp}), we obtain
 \be
 {\Delta M_{B^* B} \over {\Delta M_{D^* D}}}={m_c \over {m_b}} = {\Delta M_{B_s^* B_s} \over
 {\Delta M_{D_s^* D_s}}},
 \en
whatever the values of light quark masses ($m_u$, $m_d$, $m_s$) and other parameters appearing in any
phenomenological model are. Experimentally, the ratio of hyperfine mass splitting is given by
\cite{PDG06}
 \be
   {\Delta M_{B^* B} \over {\Delta M_{D^* D}}}\Bigg{|}_{\text{expt}} &=&
   {45.78\pm 0.35 \over{141.38\pm 0.12}}=0.3238 \pm 0.0028,  \label{bdratio} \\
   {\Delta M_{B^*_s B_s} \over {\Delta M_{D^*_s D_s}}}\Bigg{|}_{\text{expt}} &=&
   {46.1\pm 1.5 \over {143.9\pm 0.4}}=0.3204 \pm
   0.0113, \label{bdsratio}
 \en
where we take the masses of $D^*$ and $D$ mesons to the average ones
of their charged and neutral mesons. This agreement is not only a
triumph of HQET, but also reveals that the $1/m_Q$ corrections are
enough here. In addition, from the experimental data shown in Eqs.
(\ref{bdratio}) and (\ref{bdsratio}), we also find that
 \be
 {\lambda_{2s}-\lambda_{2\tilde{q}}\over {\lambda_{2s}+\lambda_{2\tilde{q}}}}\Bigg{|}_{D_sD}&=&(0.88 \pm
 0.15)~\%, \\
 {\lambda_{2s}-\lambda_{2\tilde{q}}\over {\lambda_{2s}+\lambda_{2\tilde{q}}}}\Bigg{|}_{B_sB}&=&(0.35 \pm
 1.67)~\%
 \en
This means the $SU(3)$ breaking effect of the hyperfine mass
splitting in heavy mesons is very small.

Next we consider the numerical results of mass difference between
the heavy baryons ${\cal B}^*_6$ and ${\cal B}_6$. From Eq.
(\ref{Sigma6s6}) and the ratio in Eq. (\ref{bdratio}), we predict
 \be
 \Delta M_{\Sigma^*_b \Sigma_b} = {m_c \over
 {m_b}}\Delta M^{\text{expt}}_{\Sigma^*_c \Sigma_c}= 20.9 \pm 1.0~{\text {MeV}},
 \label{predictionSigma}
 \en
where 
 \be
 \Delta M^{\text{expt}}_{\Sigma^*_c \Sigma_c}=64.4\pm 2.4~{\text {MeV}}
 \label{expSigma}
 \en
The result in Eq. (\ref{predictionSigma}) is in agreement with the
experimental data \cite{CDF}: $\Delta M^{\text{expt}}_{\Sigma^*_b
\Sigma_b} = 21.5 \pm 2.0~{\text {MeV}}$. This provides a strong vote
of confidence for the predictions of the other hyperfine mass
differences. From Eq. (\ref{Xi6s6}) and the ratio in Eq.
(\ref{bdratio}), we predict
 \be
 \Delta M_{\Xi^*_b \Xi'_b} = {m_c \over
 {m_b}}\Delta M^{\text{expt}}_{\Xi^*_c \Xi'_c}= 22.5 \pm 1.3~{\text {MeV}},
 \label{predictionXi}
 \en
where 
 \be
 \Delta M^{\text{expt}}_{\Xi^*_c \Xi'_c}=69.5 \pm 3.3~{\text {MeV}}
 \label{expXi}
 \en
From Eq. (\ref{Omega6s6}) and the ratio in Eq. (\ref{bdratio}), we
also predict
 \be
 \Delta M_{\Omega^*_b \Omega_b} = {m_c \over
 {m_b}}\Delta M^{\text{expt}}_{\Omega^*_c \Omega_c}= 22.9 \pm 0.7~{\text {MeV}},\label{predictionOmega}
 \en
where the value
 \be
 \Delta M^{\text{expt}}_{\Omega^*_c \Omega_c}=70.8\pm 1.5~{\text {MeV}} \label{expOmega}
 \en
is taken from Ref. \cite{PDG06}. In addition, from Eqs.
(\ref{expSigma}), (\ref{expXi}), and (\ref{expOmega}), we obtain
 \be
 {{\lambda^{s\tilde{q}}_2-\lambda^{\tilde{q}_1\tilde{q}_2}_2}\over {\lambda^{s\tilde{q}}_2+
 \lambda^{\tilde{q}_1\tilde{q}_2}_2}}&=&(3.8 \pm 3.1)
 \%, \label{errorsq} \\
 {{\lambda^{ss}_2-\lambda^{s\tilde{q}}_2}\over {\lambda^{ss}_2+
 \lambda^{s\tilde{q}}_2}}&=&(0.9\pm 2.6)~\%. \label{errorss}
 \en
These results reveal that the flavor $SU(3)$ breaking effect of the
hyperfine mass splitting is very small in heavy baryons, as well as
in heavy mesons.

Finally we consider the numerical results of mass difference which
is related to the heavy baryons ${\cal B}_{\bar 3}$. Combining the
experimental values \cite{PDG06,CDF,CDFLambda} and the theoretical
evaluation of $m_{\Sigma_b^0}$ \cite{hwcw}, we have
 \be
 \Delta M^{\text{expt}}_{\Sigma_c^+ \Lambda_c^+} &=& 166.4 \pm 0.4~{\text {MeV}}, \non \\
 \Delta M_{\Sigma_b^0 \Lambda_b^0} &=& 191.8 \pm 2.0~{\text {MeV}}. \non
 \en
Then we get from Eq. (\ref{Sigmaspinave})
 \be
   \delta{\bar \Lambda}^{\tilde{q}_1\tilde{q}_2}_{\Sigma_c\Lambda_c} &=&
               209.3\pm 1.6~{\text{MeV}} \non \\
   \delta{\bar \Lambda}^{\tilde{q}_1\tilde{q}_2}_{\Sigma_b\Lambda_b} &=&
               206.1\pm 2.2~{\text{MeV}}. \non
 \en
and
 \be
 {\delta{\bar \Lambda}^{\tilde{q}_1\tilde{q}_2}_{\Sigma_c\Lambda_c}-
 \delta{\bar \Lambda}^{\tilde{q}_1\tilde{q}_2}_{\Sigma_b\Lambda_b}
 \over{\delta{\bar
 \Lambda}^{\tilde{q}_1\tilde{q}_2}_{\Sigma_c\Lambda_c}+
 \delta{\bar
 \Lambda}^{\tilde{q}_1\tilde{q}_2}_{\Sigma_b\Lambda_b}}}=(0.77\pm0.65)~\%
 \en
This result reveal that, as mention above, $\delta{\bar
\Lambda}^{\tilde{q}_1\tilde{q}_2}$ is just related to the light
degrees of freedom and independent of the heavy quark flavors. Now
we use the above argument to evaluate $\delta{\bar
\Lambda}^{s\tilde{q}}$. From the data $\Delta
M_{\Xi^*_c\Xi_c}^{\text {expt}}=176.9\pm 0.9$ MeV, we obtain
 \be
 \delta{\bar \Lambda}^{s\tilde{q}}=154.4 ^{+3.8}_{-1.6} ~{\text
 {MeV}} ~~~~{\rm for}~~ \Xi^*_c \Xi_c ~~{\rm system}. \label{Lambdasq}
 \en
This result can be use to the bottomed sector due to it is also independent of the heavy flavor.
Thus, we predict
 \be
 \Delta M_{\Xi'_b \Xi_b} = 139.8^{+3.8}_{-2.0} ~{\text
 {MeV}},~~~~~ \Delta M_{\Xi^*_b \Xi_b} = 161.7^{+3.8}_{-2.0} ~{\text
 {MeV}}. \non
 \en
Combine the data $M_{\Xi_b}^{\text{expt}}=5792.9\pm 3.0$ MeV
\cite{CDF1}, we have
 \be
 M_{\Xi'_b}=5932.7 \pm 4.2 ~{\text
 {MeV}},~~~~~
 M_{\Xi^*_b}=5954.6 \pm 4.2 ~{\text
 {MeV}}. \label{predictXib}
 \en
Furthermore, we may use the Gell-Mann/Okubo formula to obtain the equal mass difference equations
 \be
 M_{\Xi'_Q}-M_{\Sigma_Q}&=& M_{\Omega_Q}-M_{\Xi'_Q}, \non \\
 M_{\Xi^*_Q}-M_{\Sigma^*_Q}&=&M_{\Omega^*_Q}-M_{\Xi^*_Q}. \label{GMO}
 \en
The accuracy of Eq. (\ref{GMO}) can be checked in charmed sector, the experimental data give
 \be
 M_{\Xi'_c}-M_{\Sigma_c} &=& 123.3\pm 2.1 ~{\text{MeV}},~~~ M_{\Omega_c}-M_{\Xi'_c} = 120.6 \pm
 3.3~{\text{MeV}}, \non \\
 M_{\Xi^*_c}-M_{\Sigma^*_c}&=& 128.4\pm 1.2 ~{\text {MeV}},~~~
 M_{\Omega^*_c}-M_{\Xi^*_c} = 121.9 \pm 3.1 ~{\text {MeV}}, \non
 \en
where the values of $M_{\Xi_c^{',*}}$ and $M_{\Sigma_c^{(*)}}$ are taken from the average masses in
charged and neutral cases. Therefore, we have the confidence to use Eq. (\ref{GMO}) in bottomed
sector. From Eqs. (\ref{predictionOmega}) and (\ref{predictXib}), we predict the masses of $\Omega_b$
and $\Omega^*_b$
 \be
 M_{\Omega_b}=6053.9 \pm 8.9 ~{\text
 {MeV}}, ~~~~~M_{\Omega^*_b}=6076.5 \pm 9.0 ~{\text
 {MeV}}.
 \en
We summarize the predictions of this work and list the other theoretical calculations and the
experimental data in TABLE II.

\begin{table}
\caption{\label{tab2} Experimental data, the predictions of this work and the other theoretical
calculation (in units of MeV).}
\begin{ruledtabular}
\begin{tabular}{c|cccccc}
 & Experiment & This work  & \cite{Jenkins}& \cite{lattice} & \cite{RQM} & \cite{KKLR} \\ \hline
 $\Delta M_{\Sigma_c^*\Sigma_c}$ & $64.4\pm2.4$ & input & $79.6\pm5.3$ & $86\pm22$& &  \\
 $\Delta M_{\Xi_c^*\Xi_c}$ & $176.9\pm0.9$ & input & & & &  \\
 $\Delta M_{\Xi^*_c \Xi'_c}$ & $69.5\pm3.3$ & input & $61.1\pm3.0$& $81\pm19$&& \\
 $\Delta M_{\Omega_c^*\Omega_c}$ & $70.8\pm1.5$ & input & $42.6\pm7.3$&$74\pm16$ & &  \\
 $\Delta M_{\Sigma^*_b \Sigma_b}$ & $21.5\pm 2.0$ & $20.9\pm 1.0$ & $23.8\pm1.6$&$24^{+13}_{-12}$&& \\
 $\Delta M_{\Xi^*_b \Xi_b}$ &&$161.7^{+3.8}_{-2.0}$&&&&\\
 $\Delta M_{\Xi^*_b \Xi'_b}$ & &$21.9\pm 1.1$&$18.3\pm0.9$&$23^{+13}_{-12}$&& \\
 $\Delta M_{\Xi'_b \Xi_b}$ &&$139.8^{+3.8}_{-2.0}$&&$148^{+35}_{-29}$&& \\
 $\Delta M_{\Omega^*_b \Omega_b}$ & &$22.9\pm 0.7$&$12.8\pm2.2$&$20\pm9$&& \\
 $M_{\Xi_b}$ & $5792.9\pm3.0$& input & & & $5812$& $5786.7\pm3.0$\\
 $M_{\Xi'_b}$ & & $5932.7\pm4.2$ &   &  & $5937$ &  \\
 $M_{\Xi^*_b}$ & & $5954.6\pm4.2$  &  &  & $5963$  &   \\
 $M_{\Omega_b}$ & & $6053.9\pm8.9$ &  &  & $6065$ & $6052.1\pm5.6$ \\
 $M_{\Omega^*_b}$ & & $6076.5\pm 9.0$  &  &  & $6088$& $6082.8\pm5.6$
\end{tabular}
\end{ruledtabular}
\end{table}
\section{Conclusion}
In this paper, based on HQET, we have presented a formalism to describe the hyperfine mass splittings
of the heavy baryons. Furthermore, through the known experimental data in charmed sector, we
predicted the hyperfine mass differences in bottomed sector. The parameters appearing in this
analysis are the ratio $m_c/m_b$ and the residual mass of heavy baryons in the heavy quark limit
$\bar {\Lambda}^{q_1 q_2}$. On the one hand the ratio $m_c/m_b$ is fixed by the experimental values
of heavy mesons, and on the other hand the residual mass difference $\delta \bar {\Lambda}^{q_1
q_2}$, due to it is independent of heavy flavor, is obtained by the known mass differences of charmed
baryons. The prediction of $\Delta M_{\Sigma^*_b\Sigma_b}$ is in agreement with the experimental
values, we expect the deviations of the other predictive mass differences are all small for the
future experimental data. In addition, in both heavy meson and baryon systems, we find that the
flavor $SU(3)$ breaking effect of the hyperfine mass splitting is very small. Finally we also
estimated the masses of $\Xi'_b$ and $\Xi_b^*$ and used the Gell-Mann/Okubo formula to calculate the
masses of $\Omega_b$ and $\Omega_b^*$. The uncertainties of these four heavy baryon masses mainly
come from the error of the measured value $M_{\Xi_b}^{\text{expt}}$. To get the more confidence in
HQET, the more precise experimental data are needed.

{\bf Acknowledgments}\\
 This work is supported in part by the National Science Council of R.O.C. under Grant No:
 NSC-96-2112-M-017-002-MY3.


\end{document}